# Valuing Student Ideas Morally, Instrumentally, and Intellectually

Amy D. Robertson

*Seattle Pacific University Department of Physics, 3307 Third Ave W, Suite 307, Seattle, WA 98119-1997*

**Abstract.** The importance of valuing student ideas in science education stands on firm empirical, theoretical, and moral grounds. However, the *reasons* for which one might value student ideas are often not explicitly distinguished, even if implicit distinctions are made in the literature. In this paper, I define and distinguish between three ways of valuing student ideas: moral, instrumental, and intellectual. I demonstrate that novice teachers in the Seattle Pacific University (SPU) Learning Assistant (LA) Program instantiate all three ways of valuing student ideas. Not only this, over the course of their participation in the LA Program, novice teachers *shift* from foregrounding the moral and instrumental value of student ideas toward highlighting their intellectual value.

## INTRODUCTION

The importance of valuing student ideas may arguably be called the "zeroth law of education research." Myriad studies have established the empirical and theoretical productivity of attending to and building on student thinking in the classroom. These practices are central to formative assessment[1] and to cognitive[2,3] and social[4,5] theories of learning.

Often, we treat "valuing" as a binary construct: one either values student ideas, or one does not. In this paper, I distinguish between three *ways* of valuing student ideas: moral, instrumental, and intellectual. Valuing students' ideas *morally* may stem from an ethic of care: one cares for students as whole people and thus values their ideas. One may also value student ideas *instrumentally,* for their usefulness in accomplishing some instructional purpose, such as diagnosing a student's misunderstanding. Or one may value student ideas *intellectually,* for their disciplinary potential or for their sensibility in light of a student's experiences. All three represent valuing student ideas but have distinct implications for how these ideas are taken up instructionally.

In the following sections, I clarify these three definitions with illustrations from novice teacher reflections on practice and connect them to relevant research literature. In addition, I document a shift in the way that a particular group of novice teachers valued student ideas: from primarily doing so morally and instrumentally toward doing so in all three ways. I present evidence for this shift in this paper, and Lovegren and I pose plausible mechanisms by which this shift came about in a second paper.[6] I argue that this shift does not represent a shift from an absence to a presence of valuing student thinking; it represents a productive shift *within* this valuing.

## METHODOLOGY

Novice teachers in the SPU LA Program (described in Lovegren and Robertson[6]) are required to submit weekly teaching reflections in which they describe the impact of pedagogy or preparatory class discussions on their teaching practice. Over the course of two quarters during the 2012-2013 academic year, I (the pedagogy/prep course instructor) noticed different ways in which LAs valued student ideas. I used LA teaching reflections to construct brief narratives that tracked the ways in which each LA valued student ideas at different points in time across the academic year. I collapsed the narratives into the three ways of valuing student ideas defined above. I then selected LA reflections that clearly and concisely illustrate the three ways of valuing.[1] In this paper, I connect these selected reflections (*cases of* the three ways of valuing I define) to the education research literature.[7,8] My goal is to contribute to the way in which the physics education research community thinks about valuing student ideas more generally.

## VALUING STUDENT IDEAS MORALLY

Novice teachers may value student ideas as part of caring for students as whole people. This valuing of ideas is *moral* in the sense that it stems from an ethic of care.[9,10] Reasons to value student ideas morally include, for example, fostering student confidence or establishing a relationship in which students feel comfortable sharing their ideas. However, there is no explicit indication that a teacher who morally values her students' ideas *does* anything with the ideas other than to affirm the student in being heard.

---

[1] Although these quotes are not necessarily representative, nor are they idiosyncratic.

LAs value student ideas morally when they notice that students are sharing their ideas (but do not explicitly attend to the substance of those ideas) and when they describe their reasons for listening to students in language consistent with an ethic of care (*e.g.,* when they want their listening to communicate care or build students' confidence). For example, one LA (Brittany[2]) writes: *"It's so refreshing [when]…[t]hey just want to share their ideas because they (hopefully) know that I care what they think!"* Likewise, Taylor writes: *"When a teacher shuts down a student's line of thinking, it can be damaging….For young students or students that lack confidence, this can be a problem."*

## VALUING STUDENT IDEAS INSTRUMENTALLY

Novice teachers may also value student ideas for their usefulness in accomplishing particular teaching and learning goals (often student achievement of the correct answer). These ideas thus become instruments in the teacher's trade. Examples include use of student ideas as instruments for diagnosing misunderstandings and/or as instructional exemplars.

### Student Ideas are Instrumental in Diagnosing Misunderstandings.

Theories of conceptual change assert that students construct new understandings on the basis of their existing ones.[11] Some researchers and practitioners interpret this to mean that effective teaching begins by understanding student ideas; this understanding is then instrumental to instructional decisions about where to begin in bringing students toward the canon.[12,13]

LAs in our courses also value student ideas as diagnostic tools, either for determining *that* students have not understood the material (a 'get-it-or-don't' conception of formative assessment[14]) or for determining *what,* specifically, the students do not understand. One LA, David, embodies the former when he describes how his diagnosis informs his decision to lead students to the correct answer: *"I began by asking probing questions about their [free body diagrams]. After a few of those, I realized that they really didn't understand the material. After that, I started to use leading questions to help guide them towards the correct response."* Another LA, Ryan, diagnoses a *specific* student misunderstanding: *"After probing for a minute [I] figured out that the student had forgotten what the cos (theta) meant in work.* [Ryan then describes a series of leading questions about the cosine of various angles and the theta-dependence of the expression for work.] *Leading through these questions she was able to see how that one variable was the determining factor in whether or not work was positive or negative."*

### Correct Student Ideas Can Serve as Instructional Exemplars.

LAs also treat correct student ideas as instrumental to other students' understandings. They describe instances in which they put examples of correct student reasoning on display, valuing their potential as tools for accomplishing the instructional goal of bringing other students toward the canon. For example, David writes: *"I…immediately noticed that one student began to provide correct responses [to my questions], while the others were lagging…behind…I asked her to explain how she came to that answer. As she walked through it, and I continued to ask questions, the other students start[ed] realizing how she got there…[O]nce I found someone who understood what was going on, I used their understanding to…spread it to the rest of the table."*

## VALUING STUDENT IDEAS INTELLECTUALLY

Ideas may also be valued intellectually – as meaningful and complex products of students' efforts to make sense of their experiences. LAs intellectually value student ideas when they treat these ideas as sensible, productive starting places for learning.[15]

Intellectual valuing is both distinct from and related to moral and instrumental valuing. Both morally and intellectually valuing student ideas include seeing student ideas for what they are, rather than what the teacher wishes them to be.[9,10] However, intellectual valuing of student ideas includes attention to the *disciplinary substance* of student ideas – the "beginnings of science" inherent in their thinking – and thus calls on extensive disciplinary knowledge.[16] Moral valuing of student ideas is part of attending to students as whole people but does not specifically attend to the substance of student thinking.

In addition, although instrumental and intellectual value for student ideas both involve *building on those*

---
[2] All names in this paper are pseudonyms.

*ideas,* they are distinct in whose meaning is at the center of the instructor's attention. For example, instantiations of intellectually valuing student ideas build more directly on and are consistent with the meaning that *students* are making of their disciplinary experiences. Instantiations of instrumentally valuing student ideas, on the other hand, often follow a logical path from the instructor's diagnosis of student thinking to the canonical answer, in many cases via a path that does not depend on the student's meaning.

## Student Ideas Are Sensible.

LAs value student ideas intellectually by making sense of canonically incorrect student reasoning, seeking to understand *why* a student may be responding the way that they are. For example, one LA, Jess, discusses a student's conflation of tension and linear mass density. She determines that the student's answer makes sense in light of the appearance of the spring: *"She…said that the spring just seemed tenser…I knew that she has interchanged the word tense to mean the tension in the spring. She thought that since spring 1 was tenser, it had more tension. This is true in the sense of how people feel tension or what we believe tension to be, but this didn't necessarily fall in line with the physics definition of tension…"*

## Student Ideas Are Productive.

LAs also illustrate intellectual value for student ideas when they treat these ideas as productive starting places for instruction. In particular, LAs describe explanations they gave that build on student ideas and experiments they designed to test student ideas. For example, Emily writes: *"One of the…misconceptions in [the Conservation of Angular Momentum] tutorial is that students do not realize that linear momentum and angular momentum are each conserved…[A few students] told me that momentum is conserved, and that therefore, some of the linear momentum turns into angular momentum, and the spinning system travels slower. The kernel [of] truth in this situation is that momentum is conserved. And since that's what the students have been taught, they applied it to the new situation. I responded by helping them revise their original statements from "momentum is conserved" to "linear momentum is conserved, and angular momentum is conserved." It was much easier to take the step from the foundation that they already had than it would have been to restart from scratch."* Emily makes sense of specific students' meaning of 'conservation of momentum' in light of her understanding of common student ideas. She builds on this meaning by distinguishing between the two different conservation principles at play.

Another LA, Sarah, designs an experiment to test a student's idea about the reflection of pulses from a fixed end: *"This week…a student was struggling with how a fixed end of a wave works…[S]he explained that her idea was that the wave would go through [the] fixed point completely leading to a flat line and then it would return. We tested it, but it never flat lined which disproved her hypothesis. She was frustrated that she had no further ideas so I pulled out a kernel of truth, that the wave would go through and return, and suggested that the portion of the wave that goes through returns leading to superposition."* The experiment that Sarah proposes tests the idea *that the student offers*. When the student realizes that her idea is incorrect, Sarah proposes an alternative explanation that connects to the student's original idea; she highlights that the wave *does* go through and return, but it begins to return as soon as the wave contacts the fixed end of the spring.

## SHIFTING TO A MORE EXPANSIVE VIEW OF STUDENT IDEAS

Over the course of Winter 2013, I noticed a shift in the way in which LAs valued student ideas, both in class and in their teaching reflections. In particular, LAs shifted from foregrounding the moral and instrumental value of student ideas (Fall 2012/early Winter 2013) to highlighting their intellectual value (mid- to late Winter 2013). Evidence for this shift is captured by the quotes I use in this paper: all of the quotes in the "Valuing Student Ideas Morally" and "Instrumentally" were written in Fall 2012 and early Winter 2013, whereas all of the quotes in "Valuing Student Ideas Intellectually" come from reflections written in mid- to late-Winter 2013. Although not every LA's reflections demonstrate this shift, many did.

This shift was not *from* a view of student ideas as value-less to a view of student ideas as valuable; it was a shift *within* LAs' ways of valuing toward a more expansive view. Nor was the shift stable; LAs often embodied all three ways of valuing student ideas.

The shift described above co-occurred with shifts in the focus of LA's teaching reflections: over the course of Winter 2013, LAs described student ideas in greater detail, and they shifted their attention from their own actions to student thinking. For example, in early Fall 2012, Emily wrote: *"I realized that I was doing too much telling to a table, and they did not really understand it. I knew that I had to do something different, and that they had to find the answer from inside of themselves. So…I had them draw a graph of* *the question, which was causing confusion. As soon as I passed the learning into their hands, they understood it."* Later, in mid-Winter 2013, Emily wrote: *"Some students asked me to clarify what the moment of inertia was…I asked them whether the hoop or the disc won the race down the ramp the week before. They told me the disc did, so I asked why. They responded by telling me that disc was going faster down the ramp, it was rotating faster, which meant that it had more $KE_{Rotational}$. Although this statement is not true, I decided to think about why they answered like this. I realized that they were saying that since the disc was covering more ground in the same amount of time, this meant that it must have had more rotations per second. After thinking about this, I was able to ask them about the conservation of energy, and taking them down that route."* Notice that in her earlier reflection, Emily modifies her approach based on an assessment of *her teaching*: she decided that she was talking too much. In the later reflection, she makes instructional decisions based on her sense-making about *student thinking*.

The co-occurrence of these shifts – toward (1) all three ways of valuing student ideas and (2) detailed attention to student thinking – suggests a two-way causal arrow between them. For example, seeking to make sense of and build on student thinking (*i.e.,* intellectually valuing these ideas) may have precipitated increased attention to this thinking; in particular, recognizing that student ideas *are* sensible and *can* be productively built on may have fostered curiosity about these ideas. At the same time, increased attention to student ideas may have further highlighted their intellectual value.

## CONCLUSION

This paper defines and distinguishes between three different ways of valuing student ideas: moral, instrumental, and intellectual. This analysis has both theoretical and methodological implications. Theoretically, these definitions challenge a binary view of "valuing": the novice teachers featured in this paper embody different *ways* of valuing student ideas, and they shift between these ways (as opposed to shifting in and out of valuing student ideas). Although drawn from a specific group of teachers in a specific context, the connection between our case and the research literature highlights the theoretical significance of these definitions and the potential for this work to broaden readers' awareness in analogous situations.[17,18] Methodologically, the illustrations of these three definitions, taken from novice teacher reflections, highlight places that other researchers might find evidence of these views of student ideas in teacher practice.

## ACKNOWLEDGMENTS

This material is based upon work supported by the National Science Foundation under Grant No. 0822342. I am grateful for the substantive contributions of Learning Assitants B. Clarke, A. Wing, A. Frazier, C. Lovegren, E. Maaske, L. Muñoz, J. Paige, K. Rininger, H. Sabo, C. Schmarr, F. Stewart, and S. Wenzinger. I also wish to acknowledge the partnership of B. Lippitt and B. Frank. Finally, I wish to thank C. Alvarado, A.R. Daane, L.S. DeWater, K.E. Gray, S.B. McKagan, J. Richards, R.E. Scherr, L. Seeley, and S. Vokos for their thoughtful feedback.